\documentstyle[aps,prl,floats,twocolumn,epsfig]{revtex}

\begin{document} 

\wideabs{
\title{Effect of Nyquist noise on the Nyquist dephasing rate in 2d
  electron systems}

\author{P.J. Burke\cite{byline}}

\address{Condensed Matter Physics, Caltech, Pasadena, CA  91125}

\author{L.N. Pfeiffer, K.W. West}

\address{Bell Laboratories, Lucent Technologies, Murray Hill, NJ 07974}
\date{\today} 
\maketitle

\begin{abstract}

We measure the effect of externally applied broadband Nyquist 
noise on the intrinsic Nyquist dephasing rate 
of electrons in a two-dimensional electron gas at low temperatures. 
Within the measurement error, the phase coherence time 
is unaffected by the externally applied Nyquist noise, including 
applied noise temperatures of up to 300~K. The amplitude of the
applied Nyquist noise from 100~MHz to 10~GHz is quantitatively 
determined in the same experiment using a microwave network analyzer.

\end{abstract}

}

What is the mechanism of electronic decoherence in disordered 2d
conductors?  One mechanism is the so-called Nyquist mechanism of 
electron-electron interactions involving small energy transfer.  
This mechanism is believed to be equivalent to the interaction of 
an electron with the flucuating electromagnetic field (i.e. the
Nyquist/Johnson noise) produced by 
all the other electrons in the system\cite{Altshuler:1985},
hence the name Nyquist dephasing. 
If this physical picture is correct, then applying a fluctuating
electric field (i.e. Nyquist/Johnson noise) from an external circuit 
should effect the coherence
time measured by weak localization in the same way as the fluctuating
electric field produced by the sample itself.  Although this effect was
first discussed over 20 years ago\cite{Altshuler:1981}, it has never 
been tested experimentally\cite{Later}. We have performed this
experiment, and present the results in this paper.

\begin{figure}
\epsfig{file=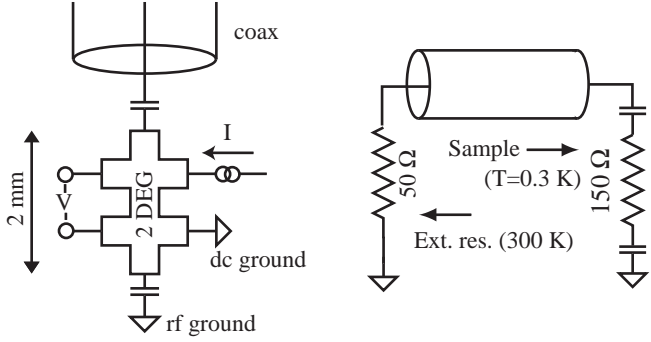}
\caption{Schematic of sample geometry.}
\label{fig:layout}
\end{figure}

The central result of this paper is to ask the following question:
What is the effect of broadband (100~MHz to 10~GHz) 
thermal fluctuations in the electric field with noise 
temperature\cite{noisetemp} 
(amplitude) of up to 300~K on the decoherence rate measured by 
weak localization on electrons in 2 dimensions?
We apply the fluctuating voltage across the terminals of a two
dimensional electron gas (see below) by terminating the room
temperature end of the coaxial cable shown in the layout in 
figure~\ref{fig:layout} with a 50~$\Omega$ resistor.
From the fluctuation-dissipation theorem, the resistor generates a noise
voltage with spectral-density given by
\begin{equation}
V_n^2=4 k_B T R,\\
\label{eq:vn}
\end{equation}
where T is the physical temperature of the external
resistor (300~K) and R is the value of the external
resistor (50~$\Omega$). The spectrum of these fluctuations is white
up to frequencies of order $k_B~T$. In the same experiment, we
measure the coupling from the sample to the resistor up to 10~GHz
with a vector network analyzer. The unique aspect of this experiment 
is that we quantify the amplitude of the applied fluctuating 
voltages to the sample terminals over a broad range of 
frequencies around $\tau_{\phi}^{-1}$ using careful microwave engineering. 
We discuss the microwave circuit models first, then
present the amplitude of the applied rf fluctuating voltage based on
these circuit models, and then discuss the measured magnetoresistance
and inferred phase-coherence times.

The sample studied is a GaAs/AlGaAs modulation-doped heterojunction grown 
by molecular beam epitaxy. The sample geometry is indicated
schematically in figure~\ref{fig:layout}. A hall bar mesa is lithographically
defined with four ohmic
contacts from diffused Ni/Au/Ge.  The sample density and mobility are
$1.25~10^{11}~cm^{-2}$ and $600,000~cm^2/V-s$, respectively, with a 
corresponding sheet resistance of roughly 80~$\Omega$/sq. 
Two additional capacitive contacts
are provided to allow for the application of high frequency
signals; these are evaporated Al gates. The
gate-2DEG separation is about 5000~$\AA$ and the gate area is about
0.25~mm$^2$, so that the capacitance value is about 50~pF. At frequencies above
roughly 100~MHz, the capacitor does not effect the rf voltage.
The d.c. current and voltage leads are several~cm long gold wires of
diameter 50~$\mu$m which act as inductive blocks at frequencies above
roughly 100~MHz. The physical temperature of the sample is held at
300~mK for the entire experiment.

\begin{figure}
\epsfig{file=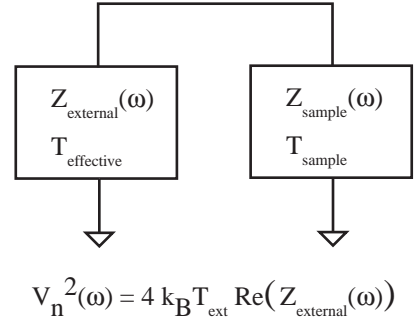}
\caption{Equivalent circuit.}
\label{fig:eqcircuit}
\end{figure}

In order to determine the amplitude of the applied fluctuating
electric field, we consider the effective circuit diagram shown in
figure~\ref{fig:eqcircuit}. The sample circuit model is that of a capacitively coupled
resistor. In reality, there will also be an inductive component at
frequencies above $\tau_{tr}^{-1}$, where $\tau_{tr}$ is the
transport (momentum) scattering time. 
(For the sample studied here, this time is
about 20~ps.) We studied this circuit model in detail in
another publication\cite{Burke:2000}, and found it to be 
valid up to 10~GHz. In this
experiment, we {\it measure}
the coupling of the sample to the coaxial cable with a microwave
vector network analyzer, as in our previous publictaion:  we measure the 
(frequency dependent) microwave reflection coefficient defined as
\begin{equation}
\Gamma(\omega)={Z_{sample}(\omega)-50~\Omega\over Z_{Sample}(\omega)+50~\Omega}.
\label{eq:gamma}
\end{equation}
This measurement is carried out by inserting the network analyzer
at the end of the coax in place of the 50~$\Omega$ resistor.
We find that the circuit model in figure~\ref{fig:eqcircuit} for the sample
describes the coupling to the sample 
(defined as $1-|\Gamma|^2$) to within 20\% over almost the
entire frequency range considered, with a sample resistance of
150~$\Omega$ and capacitance of 50 pF. 

We now consider the external circuit model. In general, since the
external resistor is ``seen'' by the sample through a coaxial cable,
the effective impedance denoted in figure~\ref{fig:eqcircuit} is a complicated function
of the frequency, length and characteristic impedance of the coax, and
the external or ``load'' resistor, with significant real
{\it and} imaginary components, even though the load resistor 
is purely real\cite{Pozar}. 
Historically, this technical difficulty has prevented quantitative
analysis of the external impedance and hence the ability to measure
its effect on phase coherence in 2d and 1d systems.
{\it However}, for the special case where the ``load'' resistance is 
equal to the characteristic impedance of the coax (50~$\Omega$ in this 
experiment), it can be shown that the effective impedance is {\it real} 
and {\it frequency independent}, and is equal to:
\begin{equation}
Z_{external}(\omega)=50~\Omega.
\label{eq:zext}
\end{equation}
In this special case, equation~\ref{eq:zext} is still 
strictly correct even if there is loss (attenuation) along the coaxial cable.

If the coax is lossless, then the equivalent circuit in figure~\ref{fig:eqcircuit} can
be used to calculate the noise voltage at the terminals of the sample, 
with the value of $T_{effective}$ given by the temperature of the
external resistor (300~K). The external circuit acts as a
noise source with voltage given by the equation in figure~\ref{fig:eqcircuit}, and a
source impedance given by $Z_{external}$.  Taking this into account, and
the fact that $Z_{external}$ is real and equal to 50~$\Omega$, the
amplitude of the voltage fluctuations at the terminals of the sample are:
\begin{eqnarray}
V_n^2(\omega)=4 k_B T_{eff} 50~\Omega~ \biggl| {Z_{sample}(\omega)\over
  50~\Omega+Z_{sample}(\omega)}\biggr|^2\nonumber\\
=k_B T_{eff} 50~\Omega~ \bigl(1-|\Gamma(\omega)|^2\bigr),
\label{eq:vneff}
\end{eqnarray}
where we have inserted the definition of $\Gamma$.  Since $\Gamma$ is
measured, we know the amplitude of the fluctuating field at each frequency.

If there is loss in the coax, then the voltage fluctuations generated
by the external resistor will get attenuated, while the coax itself
will generate some noise. By modeling the loss as uniformly
distributed along the length of the coax, and by modeling the
temperature profile along the length of the coax as linear, we find that
equation~\ref{eq:vneff} is still valid provided that the following
expression for $T_{effective}$ be used: 
\begin{equation}
T_{effective}(K) = 300 {4.3\over L}  \bigl(1- 10^{-L/10} \bigr ).
\label{eq:Teff}
\end{equation}
Here L is the loss of the entire coax, in units of dB. (0 dB is no
loss; $+~\infty$~dB is infinite loss, i.e. no transmission.)
In the limit of a lossless coax, the effective external temperature is
300~K, as expected. A more
complicated version of equation~\ref{eq:Teff} is used to calculate the
voltage noise for a 77~K ``load'', i.e. external resistor.

\begin{figure}
\epsfig{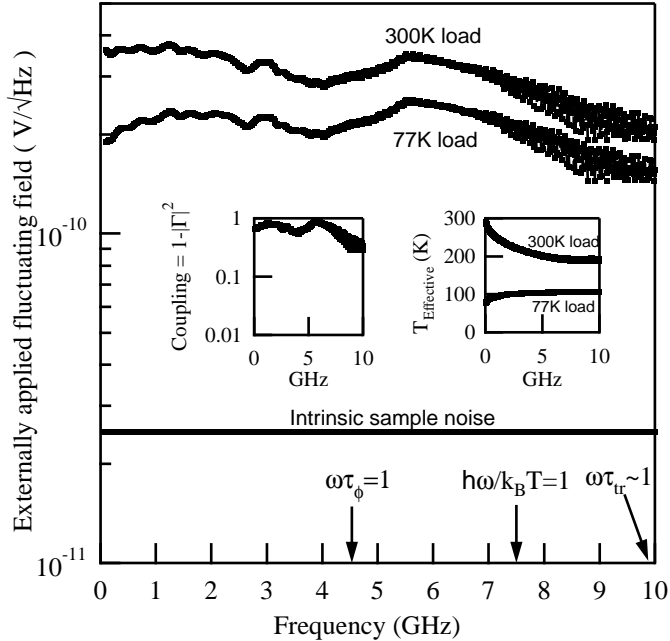}
\caption{Applied fluctuating voltage.}
\label{fig:vnoise}
\end{figure}

Finally, we plot in figure~\ref{fig:vnoise} the applied fluctuating electric field
determined from the effective external 
temperature calculated from equation~\ref{eq:vneff}
and~\ref{eq:Teff},  the measured coax
loss, and the measured sample to coax coupling efficiency. (The
measured coax loss varied from 0~dB at low frequencies to 4~dB at
10~GHz.)  The effective external temperature determined from
equation~\ref{eq:Teff} as well as the coupling measured with a network analyzer are shown
in the insets for reference. We also plot the intrinsic noise
generated by the sample itself, which is 0.3~K. In the absence of an 
external circuit, as discussed above, this is expected to be the 
dominant source of decoherence for the electrons, due to the 
so-called Nyquist dephasing mechanism.

\begin{figure}
\epsfig{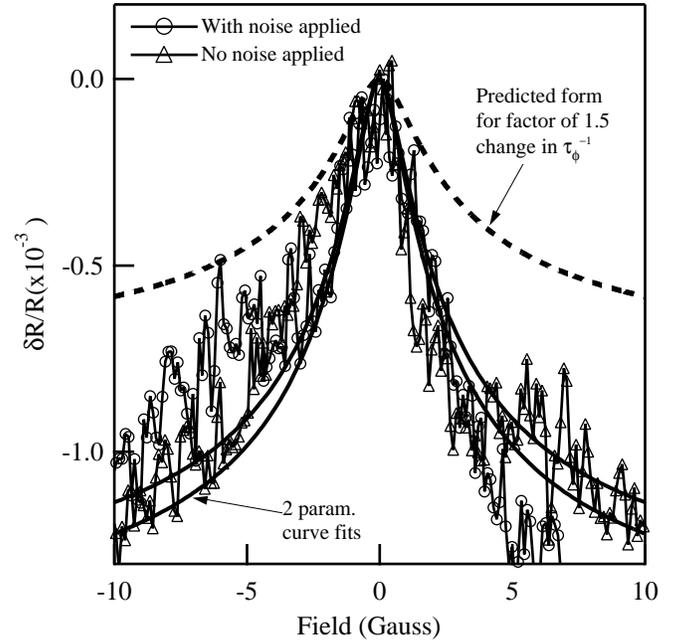}
\caption{Weak localization curve.}
\label{fig:wl}
\end{figure}

We now turn to our measurements of the phase coherence times under the
applied fluctuating fields shown in fig. 3.
Although sample resistances close to 50~$\Omega$ allow for good
characterization of the microwave coupling, they make 
measurement of weak localization difficult because of the small
resistance changes that must be resolved. Small probe
currents must be used to avoid sample heating.
For that reason the measured magnetoresistance data is somewhat noisy.
We plot in figure~\ref{fig:wl} the measured magnetoresistance for two cases: when
300~K external resistor is applied and when no external resistor is
applied. (The latter was a separate cooldown where the coax connector
was disconnected from the sample. The end of the coax was plugged to
prevent stray radiation couping to the sample from the end of the coax.)
The slight asymmetry is due to mild magnetic properties of the coaxial
connector; the magnetoresistance traces measured on the same sample in a
different mount with no connector were symmetric.
It is clear that there is no major change in the width of the peak,
hence no major change in the phase coherence time.

In order to be more quantitative, we perform a least squares fit of the
peak to the following functional 
form\cite{Hikami:1980,Dresselhaus:1992,strictly}:
\begin{eqnarray}
{\delta R \over R}={e^2 R_{s}\over \pi h}\Biggl[ 
\psi\biggl({1\over 2}+{H_{tr}\over H}\biggr)
+{1\over 2}\psi\biggl({1\over 2}+{H_{\phi}\over H}\biggr)\nonumber\\
-{3\over 2}\psi\biggl({1\over 2}+{(H_{\phi}+H_{so})\over H}\biggr)
\Biggr],
\label{eq:droverr}
\end{eqnarray}
where H is the applied magnetic field, $\psi$ is the digamma function,
and $H_i=\hbar/4 e L_i^2$, where i represents the scattering
mechanism, and $L_i=\sqrt{D \tau_i}$ the corresponding length. The i's
correspond to tr=transport, so=spin orbit, and $\phi$=phase breaking.
The elastic mean-free path and $R_{sq}$ are related, so that
there are effectively three free parameters in the theory
curve. In performing a two-parameter fit (holding $\tau_{so}$ fixed),
we find the fit results of $\tau_{\phi}$ and
$\tau_{tr}$ to be
independent of the spin-orbit scattering time, as long as $\tau_{so}$
is sufficiently larger than $\tau_{\phi}$ and $\tau_{tr}$. This is 
consistent with the results of Dresselhaus\cite{Dresselhaus:1992}, 
who studied the spin-orbit scattering rates in GaAs
2DEGs in detail. In figure~\ref{fig:wl}, we plot the fitted results for a
two-parameter fit, keeping $H_{so}$ fixed at 0.013~Gauss, the value
predicted by the Dresselhaus data for our density.  
We find a value of 34~ps and 37~ps for $\tau_{\phi}$
in the presence and absence of
the externally applied Nyquist noise, respectively\cite{ChoiNote}. 
(We find a value of 13~ps for $\tau_{tr}$ in both cases, 
in reasonable agreement with that value of 22~ps calculated 
from the measured value of $R_{sq}$.) The value of 
$\tau_{\phi}$ cannot be said to have changed within the measurement error.

From the data shown in figure~\ref{fig:wl} we can 
estimate that $\tau_{\phi}^{-1}$ changed by no more than 50\%. To
illustrate this point, we plot the
predicted curve for a factor of 1.5 change (increase) in
$\tau_{\phi}^{-1}$; this change is clearly ruled out by the experiment. 
The same conclusion applies if we perform a three-parameter fit
(varying $\tau_{so}$, $\tau_{\phi}$, and $\tau_{tr}$) or a one
parameter fit (varying only $\tau_{\phi}$ and using estimated values for
$\tau_{tr}$ and $\tau_{so}$). Thus, the experimental conclusion is
robust and independent of the particular curve-fitting procedure used.
We measure the magneto-resistance when the physical temperature
of the external resistor is changed from 300~K to 77~K, and find a 
a similar lack of change in $\tau_{\phi}$ with the change in applied
noise voltages. To an experimental
resolution of $0.1~e^2/h$, we also find no change in the B=0 conductance under 
these changes in externally applied noise.

As other groups have done\cite{Later}, we have been able to
suppress the weak localization peak by applying a constant amplitude,
single frequency field at various frequencies between 50~MHz and
20~GHz. However, we
were unable to separate the effect of heating from the electric field
induced decoherence.  It is unclear to us (theoretically) whether the
effect of a broadband, fluctuating electric field is equivalent to
that of a ``comb'' of fields with constant amplitude distributed
broadly in frequency. Even if this were predicted theoretically, it
is still important to test experimentally. The case of an externally applied
broadband fluctuating field is in some sense a better check of the
theory of the Nyquist dephasing mechanism, since the fields generated by the
electrons in the sample are themselves broadband and fluctuating.

For the sake of comparison with other experiments which measure the
effect of a single frequency constant amplitude field, we can estimate
the (rms) value of the electric field strength in our experiment. 
For single frequency constant amplitude experiments, the important 
dimensionless measure of the field strength is given by
$\alpha = 2 e^2 D E^2/[\hbar^2(2 \pi f)^3]$, with D the diffusion
constant and E the electric field strength. In our experiments, we
calculate that the rms voltage is roughly 10~mV, hence an rms electic
field strenght of 10 V/m. We estimate that $\alpha\approx~4$. For
fixed amplitude and frequency experiments, this
is theoretically\cite{Altshuler:1981} enough amplitude 
to change $\tau_{\phi}$ by 100\%, if
a single frequency constant amplitude with the same field strength
were applied.

We now turn to the theoretical interpretation of our results. There
have only been a few theoretical calculations of the effect of
external circuit noise on the phase-coherence of electrons in 2d
conductors\cite{Altshuler:QTS1982,Altshuler:JPC1982}.
Equation 3.3.22 of reference\cite{Altshuler:QTS1982} seems to predict
that in the presence of external circuit noise with 
noise temperature $T_0$\cite{noisetemp}, and with good circuit
coupling as we have in this experiment,
the measured $\tau_{\phi}$ should be comparable to the value that one would
measure in the absence of such noise if the physical temperature of
the electrons was equal to $T_0$. For the experiment considered here,
that would imply that our measured value of $\tau_{\phi}$ in the
presence of 300~K of noise should be essentially zero, fully
suppressing the weak localization peak, in
contradiction to what we observe. For these reasons,
we are perplexed as to why the weak localization peak still exists at all,
even in the presence of such a high artificial temperature of the
electromagnetic field fluctuations. 

We speculate on three possible
reasons for this robustness of the phase coherence to the externally
applied noise.  First, it could be due to the fact that in our sample,
the electron motion is ballistic at frequencies above
$\tau_{tr}^{-1}$, 
which is comparable to $\tau_{\phi}^{-1}$ in this experiment.
Second, even though the amplitude of the field is up to
300K, its frequency is probably not coupled to the sample all the way
up to 300~K/$k_B$, which is many THz.  There are, to our knowledge, no
theoretical predictions of what happens to $\tau_{\phi}^{-1}$ 
if a thermal field
is applied only over a certain (albeit broad) range of frequencies,
corresponding to the situation in our experiment.  Finally, we
speculate that perhaps in order to efficiently cause dephasing, 
not only must the frequency of the applied thermally
fluctuating field by comparable to $\tau_{\phi}^{-1}$, but that 
the wave vector may also need to be of order
$L_{\phi}^{-1}$. In our experiments, the wave vector of the applied
fluctuating field is roughly 1/(1~mm), which is
much smaller than $1/L_{\phi}$, roughly 1/(10~$\mu$m).  These hypotheses
suggest a class of future experiments to determine the effect of
fluctuating electric fields on systems with shorter mean free paths 
(such as thin films), as well as other coupling geometries to allow 
the coupling of higher-wavevector fluctuating electric fields. 

In conclusion, we have measured the effect of externally applied
broadband Nyquist noise on the intrinsic Nyquist dephasing rate of
electrons in 2 dimensions.  Within the experimental accuracy, the
phase coherence time is unaffected by the externally applied Nyquist
noise, including applied noise temperatures of up to 300~K.

It is a pleasure to acknowledge J.P. Eisenstein for his
insightful expertise, advice, and support. This work was supported by 
Sandia National Labs under Grant No. DE-AC04-AL85000 
and the DOE under Grant No. DE-FG03-99ER45766. One of the authors (P.J.B.) 
was supported in part by the Sherman Fairchild Foundation.


\begin{references}
\bibitem{noisetemp}
The noise temperature is a standard measure of the amplitude of voltage
  fluctuations, defined such that the spectral density of voltage fluctuations
  corresponding to a noise temperature T is given by $k_B T R$, with R equal to
  50~$\Omega$. This is a convenient unit since, according to the
  fluctuation-dissipation theorem, the noise temperature of the Johnson/Nyquist
  noise produced by a 50~$\Omega$ resistor is equal to its physical
  temperature.
\bibitem{strictly}
Strictly speaking, equation 6 is valid only at fields smaller than $H_{tr}$,
  which for the sample measured here is $\sim$~1~Gauss. We assume this curve
  holds semi-quantitatively for larger values of H in order to get a
  semi-quantitative estimate of $\tau_{\phi}$.
\end{references}

\end{document}